\documentclass[apj]{emulateapj}   
\usepackage{graphicx}
\usepackage{multirow}

\usepackage{amsmath}
\usepackage{natbib}
\usepackage{hyperref}
\usepackage[usenames,dvipsnames]{color}  
\usepackage[title]{appendix}
\usepackage[normalem]{ulem}
\usepackage[dvipsnames]{xcolor}

\def\msun{{\rm ~M}_{\odot}}
\def\rsun{{\rm ~R}_{\odot}}

\def\gpy{{\rm ~Gpc}^{-3} {\rm ~yr}^{-1}}
\def\kms{{\rm ~km} {\rm ~s}^{-1}}
\def\mpy{{\rm ~M}_{\odot} {\rm ~yr}^{-1}}

\def\erg{{\rm ~erg}}

\begin{document}

\title{The Uncertain Future of Massive Binaries Obscures the Origin of LIGO/Virgo Sources}

\author{
   K. Belczynski\altaffilmark{1}, A. Romagnolo\altaffilmark{1}, A. Olejak\altaffilmark{1}, 
   J. Klencki\altaffilmark{2}, D. Chattopadhyay\altaffilmark{3}, S. Stevenson\altaffilmark{3}, 
   M.~Coleman~Miller\altaffilmark{4}, J.-P. Lasota\altaffilmark{1,5}, 
   Paul A. Crowther\altaffilmark{6}
}

\affil{
   $^{1}$ Nicolaus Copernicus Astronomical Center, Polish Academy of Sciences,
          ul. Bartycka 18, 00-716 Warsaw, Poland (chrisbelczynski@gmail.com,
          aleksandra.olejak@wp.pl, amedeoromagnolo@gmail.com)\\
   $^{2}$ Department of Astrophysics/IMAPP, Radboud University, P.O. Box 9010, 6500 GL
          Nijmegen, The Netherlands (kklencki@gmail.com)\\
   $^{3}$ Centre for Astrophysics and Supercomputing, Swinburne University of Technology, John St., 
          Hawthorn, Victoria- 3122, Australia (dchattopadhyay@swin.edu.au, 
          simonpaulstevenson@gmail.com)\\
   $^{4}$ Department of Astronomy and Joint Space-Science Institute University 
          of Maryland, College Park, MD 20742-2421, USA
          (miller@astro.umd.edu)\\
   $^{5}$ Institut d'Astrophysique de Paris, CNRS et Sorbonne Universit\'e,
          UMR 7095, 98bis Boulevard Arago, 75014 Paris, France
          (lasota@iap.fr)\\
   $^{6}$ Department of Physics \& Astronomy, University of Sheffield, Hounsfield Rd, Sheffield, S3 7RH, 
          UK (paul.crowther@sheffield.ac.uk)
}

\begin{abstract}
The LIGO/Virgo gravitational--wave observatories have detected at least 50 double black hole (BH) 
coalescences. This sample is large enough to have allowed several recent studies to draw conclusions
about the implied branching ratios between isolated binaries versus dense stellar clusters as the 
origin of double BHs. It has also led to the exciting suggestion that the population is highly likely 
to contain primordial black holes. Here we demonstrate that such conclusions cannot yet
be robust, because of the large current uncertainties in several key aspects of binary stellar 
evolution. These include the development and survival of a common envelope, the mass  and angular 
momentum loss during binary interactions, mixing in stellar interiors, pair-instability mass loss and 
supernova outbursts. Using standard tools such as the rapid population synthesis codes {\tt StarTrack} 
and {\tt COMPAS} and the detailed stellar evolution code {\tt MESA}, we examine as a case study the 
possible future evolution of Melnick 34, the most massive known binary star system (with initial
component masses of $144\msun$ and $131\msun$). We show that, despite its fairly well-known orbital 
architecture, various assumptions regarding stellar and binary physics predict a wide variety of 
outcomes: from a close BH-BH binary (which would lead to a potentially detectable coalescence), through 
a wide BH-BH binary (which might be seen in microlensing observations), or a Thorne-{\.Z}ytkow object, 
to a complete disruption of both objects by pair-instability supernovae. Thus since the future of 
massive binaries is inherently uncertain, sound  predictions about the properties of BH-BH systems are 
highly challenging at this time. Consequently, drawing conclusions about the formation channels 
for the LIGO/Virgo BH-BH merger population is premature.
\end{abstract}

\keywords{stars: black holes, neutron stars, x-ray binaries}

\section{Introduction}
\label{sec.intro}

The LIGO/Virgo Collaboration (LVC) has reported gravitational-wave detections of $\sim 50$ double 
black hole (BH-BH) coalescences~\citep{Abbott2021a}. The majority of these can be explained as 
originating through any of several channels, including isolated binary evolution, dynamics in dense 
stellar clusters, or primordial black holes~\citep{Mandel2021a}. Until now, observations contain only 
hints of the possible origin of observed double BHs. For example, as anticipated prior to the 
detections~\citep{Belczynski2010a}, many of the black holes in these binaries have masses 
$\sim 30\msun$ or larger, which is considerably in excess of the most massive stellar-origin BHs 
known through electromagnetic observations. Another trend is that the effective spin of the binaries 
(which is the mass-weighted projection of the black hole spins onto the orbital axis) is low; this 
could be an indication of random orbits from dynamical processes, or could point toward intrinsically 
low spins produced by efficient angular momentum transport in massive stars \citep{Spruit2002,Farr2017,
Vitale2017b,2018ApJ...854L...9F,Fuller2019b,Bavera2020,Belczynski2020b}.

However, there are also individual events with characteristics that may be more challenging to 
explain, and which therefore hold promise for discriminating between formation channels. One such 
event is GW190814~\citep{gw190814}, which is an extremely asymmetric binary consisting of a 
$\approx 23\msun$ black hole and a $2.6\msun$ object that is either the lightest black hole or the 
heaviest neutron star yet detected. Another is the double black hole event GW190521~\citep{gw190521a}, 
which has two black holes which may have masses of $\sim 85\msun$ and $\sim 65\msun$, putting them 
both in the pair-instability mass gap, although it is possible that the black hole mass ratio is 
farther from unity and both black holes avoid the gap \citep{Fishbach2020c,2021ApJ...907L...9N}.

Based on these results, several groups have recently analyzed the BH-BH population as a whole, with 
special attention to outliers such as GW190814 and GW190521, to obtain insight into the relative 
fraction of events from different formation channels. For example, \cite{Zevin2021} studied a 
mixture of isolated binary evolution and dynamical formation in globular clusters and concluded that 
neither channel can contribute more than $70\%$ to the LIGO/Virgo observed population of BH-BH 
mergers. In contrast, \cite{Ossowski2021} disfavored the globular cluster channel in favor of 
isolated binary evolution. \cite{Franciolini2021} investigated four formation channels (isolated 
binaries, globular clusters, nuclear star clusters, and primordial BHs) and found a high likelihood 
that primordial BH-BH mergers are part of LIGO/Virgo source population. 

All of these studies perform proper model comparison to infer which model or mixture of models is 
favored. However, model comparison requires precisely specified models, i.e. models with precisely 
defined physics. Here we emphasize that the physics uncertainties~\citep{Schootemeijer2019} in at 
least one of those models, of isolated binary evolution, are sufficiently large that (to put it in 
Bayesian terms) the prior dominates the conclusion. That is, different assumptions can lead to very 
different outcomes, which means that population studies are not yet at the stage allowing strong and 
credible  conclusions to be drawn. 

We demonstrate these model weaknesses using various, different assumptions in performing simulations 
whose aim is to determine the fate of the most massive known binary system. Despite the high binary 
frequency of massive stars \citep{Sana2012}, eclipsing systems  with primary masses $\gg 50 \msun$ 
are exceptionally rare. Within the Milky Way the most extreme double-lined  systems are located in 
young, rich star clusters: A1 within NGC~3603 \citep{Schnurr2008}, F2 within the Arches 
\citep{Lohr2018} and WR20a in Westerlund 2 \citep{Bonanos2004}. All three are short period (days), 
low eccentricity systems with main sequence Wolf-Rayet (H-rich, WNh) primaries. 
 
The most massive double-lined eclipsing system in the  Large Magellanic Cloud (LMC) is also a short 
period system, Hunter 38 in the Tarantula Nebula with a O-type primary, whose mass is $\sim 57\msun$ 
\citep{Massey2002}. The most extreme LMC binaries are non eclipsing systems, also within the Tarantula 
Nebula. They have minimum dynamical primary  masses in excess of $\sim 50 \msun$, albeit with orbits 
whose periods are an order of magnitude longer and high eccentricities. From comparison with Bonn 
stellar evolutionary models at LMC metallicity \citep{Brott2011},  R139 ($P_{\rm orb}= 154$\,d, 
$e=0.38$) has a primary O supergiant mass of $\sim 80 \msun$ \citep{Mahy2020}, R144 ($P_{\rm orb}= 74$\,d, 
$e=0.51$)  has a primary WNh mass of $\sim 110 \msun$ \citep{Shenar2021} and Melnick~34 (Mk~34, 
$P_{\rm orb}=155$\,d, $e=0.68$) -- the current record-holder - has component WNh+WNh masses of 
$139^{+21}_{-18} \msun$ and $127\pm17 \msun$ \citep{Tehrani2019} with initial masses of $144\msun$ 
and $131\msun$. \citet{Pollock2018} first established that Mk~34 is a colliding wind binary from 
analysis of X-ray time series observations, while  \citet{Tehrani2019} noted that the potential fate 
of Mk~34 involves a double stellar mass black hole binary merger.

We predict the fate of Mk~34, using different physics assumptions, applying two rapid population 
synthesis codes ({\tt StarTrack} and {\tt COMPAS}) and the detailed stellar evolutionary code 
({\tt MESA}). We find a wide variety of possible outcomes (see Tab.~\ref{tab.models}) in terms of 
the black hole masses and orbital separations, and even in terms of whether black holes will form at 
all. We therefore urge caution in drawing important and credible conclusions about the LVC BH-BH 
population based on models of massive--binary evolution containing, by necessity, uncertain physics.

\begin{table}
\caption{Fate of Mk~34}
\begin{tabular}{lcccc}
\hline\hline
model         & $M_{\rm BH1}$ & $M_{\rm BH2}$ & $t_{\rm delay}$ &  fate         \\
              &     [$\msun$] &     [$\msun$] &           [Myr] &               \\
\hline\hline
StarTrack1    &          22.5 &           22.1 &            47.5 &  close BH-BH?$^{a}$ \\    
StarTrack2    &          35.7 &           33.3 &          10,035 &  close BH-BH?$^{a}$ \\    
StarTrack3    &          36.0 &           32.5 &          11,663 &  close BH-BH?$^{a}$ \\    
              &               &                &                 &                \\
COMPAS1       &          19.8 &           20.2 &  $>t_{\rm hub}$ &     wide BH-BH \\
COMPAS2       &          31.8 &           31.7 &  $>t_{\rm hub}$ &     wide BH-BH \\
COMPAS3       &          31.8 &           31.7 &  $>t_{\rm hub}$ &     wide BH-BH \\
              &               &                &                 &                \\
MESA1         &          21.9 &     51.6$^{b}$ &            ---  & Thorne-{\.Z}ytkow \\
MESA2         &          35.2 &     80.9$^{b}$ &            ---  & Thorne-{\.Z}ytkow \\
MESA3         &          35.3 &     85.4$^{b}$ &            ---  & Thorne-{\.Z}ytkow \\
              &               &                &                 &                \\
Pavlovskii1   &          21.9 &           22.1 &  $>t_{\rm hub}$ &     wide BH-BH \\
Pavlovskii2   &          35.2 &           33.3 &  $>t_{\rm hub}$ &     wide BH-BH \\
Pavlovskii3   &           --- &            --- &             3.3 & stellar merger$^{c}$ \\
              &               &                &                 &                \\
QuasiSingle1  &           --- &            --- &             --- &  PSN+PSN$^{d}$ \\
QuasiSingle2  &     $\sim 60$ & $\sim 60\msun$ &  $>t_{\rm hub}$ &     wide BH-BH \\
QuasiSingle3  &     $\sim 30$ & $\sim 30\msun$ &  $>t_{\rm hub}$ &     wide BH-BH \\
QuasiSingle4  &     $\sim 20$ & $\sim 20\msun$ &  $>t_{\rm hub}$ &     wide BH-BH \\
\hline
\hline
\end{tabular}
\\$^{a}$: optimistic (non-standard) {\tt StarTrack} models are used to get this result
\\$^{b}$: for MESA models we list CE donor mass in column $M_{\rm BH2}$
\\$^{c}$: merger of post-MS star and MS star: formation of very massive single star, 
          fate: PSN or a single BH
\\$^{d}$: pair-instability supernovae disrupting binary components 
\label{tab.models}
\end{table}

\section{Calculations}
\label{sec.calc}

For initial properties of Mk~34 we select: $M_{\rm a}=144\msun$, $M_{\rm b}=131\msun$, $e=0.68$, 
$a=760\rsun$ chosen to result in an orbital period of $P_{\rm orb}=155$\,d after $0.6$\,Myr 
\citep[current age,][]{Tehrani2019} of system evolution with the {\tt StarTrack} code. We adopt 
the LMC metallicity of $Z=0.006$ \citep{Rolleston02}.

\subsection{StarTrack calculations}
\label{sec.startrack}

We use the population synthesis code {\tt StarTrack}~\citep{Belczynski2020b}, which employs 
analytic fits to evolutionary tracks of non-rotating stellar models~\citep{Hurley2000}.  
We adopt standard wind losses for massive stars from~\cite{Vink2001} and LBV winds as  
$(dM/dt)_{\rm lbv}=f_{\rm lbv} 10^{-4}\mpy$ with $f_{\rm lbv}=1.5$ from \cite{Belczynski2010b}. 

For stars that overfill their Roche lobes, we initiate mass transfer between binary components 
and associated (if any) mass loss from binary systems. If the binary is not circularized by tidal 
interactions we circularize it ($e_{\rm new}=0$) to periastron distance ($a_{\rm new}=a(1-e)$) 
in one timestep and only then start Roche lobe overflow (RLOF). For nuclear-timescale mass 
transfer (NTMT) and thermal-timescale mass transfer (TTMT) we use the standard formalism, while we 
use a diagnostic mass ratio diagram as a criterion for common envelope (CE) development (see 
Sec.5 of \cite{Belczynski2008a}). During the TTMT/NTMT the fraction of mass lost by the donor star that 
is accumulated by non-degenerate companion stars is set to $f_{\rm a}=0.5$, while the rest is 
lost with specific angular momentum (expressed in units of $2 \pi a^2 / P_{\rm orb}$) of 
$j_{\rm loss}=1.0$ (see eq. 33 of \cite{Belczynski2008a}). The accumulation of mass on compact 
objects (e.g., NS/BH) is limited by the (Eddington) critical accretion rate and mass is lost 
with the specific angular momentum of the compact accretor~\citep{King2001,Mondal2020}. 

We employ the delayed core-collapse supernova (SN) engine in NS/BH mass calculation
~\citep{Fryer2012} which allows for populating the lower mass gap between NSs and BHs
~\citep{Belczynski2012a,Zevin2020}. We employ weak pair-instability pulsation supernova 
(PPSN) mass-loss and pair-instability supernova (PSN) model that results in upper mass gap: 
no BHs with mass $M_{\rm BH} \gtrsim 55\msun$~\citep{Belczynski2020b}. We allow for the 
fallback decreased NS/BH natal kicks with $\sigma=265\kms$ and no natal kicks for direct BH 
formation. This is our standard input physics marked as "StarTrack1" model in 
Table~\ref{tab.models}.

The development of the CE phase is a big issue in stellar/binary astrophysics
~\citep{Ivanova2013a,Olejak2021}. We are agnostic about which systems should be sent to a CE and 
which should evolve through stable RLOF. In {\tt StarTrack} models we allow for the most optimistic 
scenario (see Sec. \ref{sec.evol1}), and we send nearly all systems through the CE to form 
(potentially) BH-BH mergers. We do not do this on regular basis. According to our standard input 
physics, donors with radiative envelopes (e.g., in the Hertzsprung gap) do not enter the CE phase. 
Since it is not fully  understood how exactly a CE develops, we test various assumptions to show 
how these influence the future fate of binary systems such as Mk~34. Contrasting models are being 
presented as well. 

During CE events the entire envelope of the donor is assumed to be lost from the binary, with the 
exception of compact object companions that are allowed to accrete a small fraction of donor's 
envelope at $5\%$ of the Bondi rate~\citep{MacLeod2017a}. The CE orbital decay is calculated 
with the standard energy-balance formalism~\citep{Webbink1984} in which we adopt a $100\%$ efficiency 
of the orbital energy transfer ($\alpha=1.0$) into the envelope, while the binding energy is 
parameterized by detailed stellar models ($\lambda$ scaling:~\cite{Xu2010,Dominik2012}). 

In ``StarTrack2" we decrease wind mass--loss rates for LBV stars to $f_{\rm lbv}=0.48$ and we 
increase the He core mass at the end of the main sequence by a factor of $f_{\rm core}=1.5$ with
respect to the original~\cite{Hurley2000} models. This model approximately reproduces  
the basic properties of the $131\msun$ and $144\msun$ models at terminal-age main sequence (TAMS)
obtained in our MESA computations (see Sec.~\ref{sec.mesa}).

In ``StarTrack3" we circularize massive binaries with angular momentum conservation 
($a_{\rm new}=a(1-e^2)$), we set $j_{\rm loss}=0.1$, and $f_{\rm a}=0.25$ while keeping the 
rest of the input physics as in ``StarTrack2". This model aims to test the survival of the CE phase in 
an Mk~34-like future evolution (see Sec.~\ref{sec.evol}). In practice, such set-up allows the 
secondary star of Mk~34 to have a large radius (wide binary orbit) during the RLOF and therefore 
potentially to develop a convective envelope but to survive the CE phase.

\subsection{COMPAS calculations}
\label{sec.compas}

We use the population synthesis code {\tt COMPAS}~\citep{Stevenson:2017tfq,Vigna-Gomez:2018dza,
Chattopadhyay:2020lff}, which incorporates stellar~\citep{Hurley2000} and binary evolution 
~\citep{Hurley2002} with updated wind prescriptions for massive stars~\citep{Vink2001,
Belczynski2010b}. The LBV wind losses are as in {\tt StarTrack} (see Sec.~\ref{sec.startrack}). 

In its default set-up, the mass accretion during RLOF onto degenerate stars (i.e. black holes, 
neutron stars and white dwarfs) is Eddington limited in {\tt COMPAS}. For non-degenerate stars, this 
accretion limit is set by the ratio of the rate at which mass is being donated to the rate of which 
mass can be incorporated into the companion star. These two rates, in turn, are determined by the 
thermal (Kelvin-Helmholtz) timescales of the donor and the companion respectively
~\citep{Kalogera:1995yc}. The thermal timescale is an explicit function of the mass, radius and 
luminosity of the star. For a star of total mass $M$, envelope mass $M_{\rm env}$, radius $R$ and 
luminosity $L$, the thermal timescale $\tau_{\rm KH}$ is given by $\tau_{\rm KH}=GMM_{\rm env}/RL$, 
where $G$ is the universal gravitational constant. In binaries, since these stellar parameters are 
determined by the orbital period at which the donor overflows its Roche lobe, the thermal timescale 
of the donor becomes an implicit function of the orbital properties of the binary \citep{Schneider2015}. 
Thus the accretion efficiency in binaries is primarily determined by their orbital period. 
The mass transfer efficiency $\beta_\mathrm{acc}$ (ratio of the mass gained by the donor to the mass 
lost from the companion)\footnote{Equivalent of the parameter $f_{\rm a}$ in {\tt StarTrack} 
described in Sec.~\ref{sec.startrack}.} in a non-degenerate star can be thus expressed as 
$\beta_\mathrm{acc}= \mathrm{min}(1, 10 \frac{\tau_\mathrm{KH,acc}}{\tau_\mathrm{KH,don}})$, where 
$\tau_\mathrm{KH,acc}$ and $\tau_\mathrm{KH,don}$ are the thermal timescales of the accretor and 
donor respectively~\citep{Hurley2002,Schneider2015,Chattopadhyay:2020lff}. 

The stability of the mass loss is determined by the parameter $\zeta$ (critical mass-radius exponent 
for development of CE) in {\tt COMPAS} using fits from the \cite{Ge2015} simulations as described in 
~\cite{Vigna-Gomez:2018dza}. In nearly equal-mass, close binaries the thermal timescales of the donor 
and accretor being very similar, the mass transfer is usually conservative and remains stable. In 
close systems with more extreme mass ratio, the thermal timescale of the donor being much longer 
than the accretor, the mass transfer becomes non-conservative leading to a CE phase. The
~\cite{Ge2015} criteria renders the mass transfer from evolved (non-main sequence, non-degenerate) 
massive stars as predominantly stable \citep{Neijssel2019}, which is very similar to the 
~\cite{Pavlovskii2017} model (discussed in section.~\ref{sec.pavlovskii}). We assume an isotropic 
re-emission model for angular momentum loss during non-conservative stable RLOF~\citep{Pols1998}.

Unlike in {\tt StarTrack}, the binary is not circularized right before RLOF~\citep{Vigna-Gomez:2018dza}. 
But binaries that survive the CE events are always circularized~\citep{Vigna-Gomez:2018dza}. All 
other default RLOF and CE mass transfer specifications in {\tt COMPAS} are identical to 
{\tt StarTrack} as described in Sec.~\ref{sec.startrack}. 
 
Our standard model COMPAS1 utilizes the~\cite{Fryer2012} ``delayed' supernovae prescription and  
pre-supernova core mass to post-supernova remnant mass mapping. The (pulsational) pair-instability 
supernovae modeling is implemented in COMPAS \citep{Stevenson2019} with polynomial fitting from the 
models by~\cite{Marchant2018} as the default input. The natal kick distributions (including 
fallback) for BHs and NSs are identical to StarTack1 model. 
 
In the model COMPAS2 we reduce the LBV wind mass loss rate to $f_\mathrm{lbv}=0.48$ (from 
$f_\mathrm{lbv}=1.5$ in COMPAS1). We also increase the He core mass of the terminal main sequence 
stars by a factor of $f_\mathrm{core}=1.5$ multiplied to the fitting formula from \cite{Hurley2000} 
(eq. 30). COMPAS2, like StarTrack2 is adjusted to approximately reproduce the total mass and the 
core mass at TAMS of $131\msun$ and $144\msun$ models computed with {\tt MESA}. 
 
The model COMPAS3 is identical to COMPAS2 but here we allow the binary to circularize (while 
conserving its angular momentum) right before the onset of RLOF as detailed for {\tt StarTrack} in 
Sec.~\ref{sec.startrack}.

\subsection{MESA calculations}
\label{sec.mesa}

\subsubsection{Calibration of TAMS core masses}

\label{sec:mesa_calib}

The fitting formula by \citet{Hurley2000} to the evolutionary tracks from \citet{Pols1998}, which 
are the basis for {\tt StarTrack} and {\tt COMPAS} population synthesis codes, are based on stellar 
models computed for stars with masses up to $40\msun$. The treatment of more massive stars in 
{\tt StarTrack} and {\tt COMPAS} relies on extrapolation. In the mass range considered in this study 
($130-145\msun$), this can lead to a significant deviation in basic stellar properties from what 
detailed stellar models produce (or what is inferred from observations). One property that is 
particularly inaccurate due to the extrapolation of \citet{Hurley2000} formulae (and which is 
crucial for considerations of the final fate of the Mk~34 system) is the ratio of the helium core 
mass to the total star mass at TAMS ($M_{\rm core,TAMS}/M_{\rm TAMS}$). For very massive stars this 
ratio is close to unity \citep{Yusof2013, Kohler2015}. In contrast, a $144\msun$ star at $Z=0.006$ 
metallicity evolved with {\tt StarTrack} or {\tt COMPAS}, even though significantly stripped through 
MS winds ($M_{\rm TAMS}=77.4\msun$) is far away from being a helium star with the helium core mass 
of only $M_{\rm core,TAMS}=34.1\msun$. 

To correct for this and calibrate the properties of the {\tt StarTrack} and {\tt COMPAS} models at TAMS, 
we use the {\tt MESA} 1D stellar-evolution code~\citep{paxton_2011_aa,paxton_2013_aa,Paxton2015,
paxton_2018_aa,paxton_2019_aa}\footnote{MESA version r15140,  \url{http://mesa.sourceforge.net/}}.
We compute single models of $131\msun$ and $144\msun$ stars at $Z=0.006$ metallicity. The relative 
initial abundances of metals follow \citet{Grevesse1996}. We model convection by using 
mixing-length theory \citep{BohmVitense1958} with a mixing-length parameter $\alpha=2.0$, and we 
adopt the Schwarzschild criterion for convection. We used the Dutch wind setup in {\tt MESA}, which 
combines different prescriptions depending on the effective temperature $T_{\rm eff}$ and the 
fractional surface hydrogen abundance $H_{\rm sur}$. As shown in Table~\ref{tab.dutch}, for 
$T_{\rm eff}<10^4$ K the code uses the mass-loss rates from \citet{deJager1988}, regardless of the 
hydrogen surface abundance. For $T_{\rm eff}\geq10^4$ K, {\tt MESA} adopts either the 
\citet{NugisLamers2000} prescriptions (if $H_{\rm sur}<0.4$), or mass-loss rates from \cite{Vink2001} 
(if $H_{\rm sur}\geq0.4$). Additionally, mass loss rates in {\tt MESA} can be increased or decreased 
by changing a specific scaling factor $f_{\rm wind}$. The standard prescription of \citet{Vink2001} 
is known to underestimate the empirical mass-loss rates of very massive MS stars, which increase 
dramatically as they approach the Eddington limit, $\Gamma_e$ 
\citep{Vink2011b,Bestenlehner2014,Bestenlehner2020a}. Indeed, clumping corrected mass-loss rates of 
the components of Mk~34 from \citet{Tehrani2019} exceed \citet{Vink2001} prescriptions by factors of 
2--3. 

\begin{table}
\caption{Dutch Stellar Winds in {\tt MESA}$^{a}$}
\begin{tabular}{ccc}
\hline\hline
   & \textbf{$T_{\rm eff}<10^4$ K} & \textbf{$T_{\rm eff}\geq10^4$ K} \\
\hline\hline
 --                     & \citet{deJager1988} & -- \\
 $H_{\rm sur}<0.4$      & -- & \cite{NugisLamers2000}  \\
 $H_{\rm sur} \geq 0.4$ & -- & \citet{Vink2001} \\
\hline
\hline
\end{tabular}
\\$^{a}$: based on $T_{\rm eff}$ winds either depend or do not depend on $H_{\rm sur}$
\label{tab.dutch}
\end{table}

We account for convective overshooting above the hydrogen-burning core by applying the step 
overshooting formalism, which extends the convective core by a fraction $\delta_{ov}$ of the local 
pressure scale height. 

We initialize our models with the initial rotation of $V_{\rm i}=250\kms$ \citep[guided by the 
analysis of][]{Tehrani2019}. For rotational mixing, we include the effects of the Eddington-Sweet 
circulation, secular shear instabilities, and the Goldreich-Schubert-Fricke instability, with an 
efficiency factor $f_{\rm c}=1/30$ \citep{Heger2000,Brott2011}.

We avoid using the MLT++ option in {\tt MESA} \citep{paxton_2013_aa}. As a result, models that reach 
the red supergiant stage encounter numerical difficulties in their superadiabatic outer envelope 
layers \citep{Pavlovskii2015,Klencki2020}, which prohibits us from following their evolution to the 
point of maximum radial expansion. However, for the purpose for the current study, we are only 
interested in the properties of {\tt MESA} models at TAMS as well as whether or not the models 
expand sufficiently to lead to a RLOF in the Mk~34 binary system. We thus stop the {\tt MESA} 
computations when the radius of $2000\rsun$ is reached. Otherwise, we stop the simulation after 
10,000 {\tt MESA} steps. Such stopping conditions are sufficient for our purposes in all the 
considered scenarios for the Mk~34 system. 

For calibration of population synthesis models at TAMS we calculate a {\tt MESA} stellar model with 
initial mass $M_{\rm ZAMS}=144\msun$, metallicity $Z=0.006$, and $V_{\rm i}=250\kms$. We apply the 
standard Dutch winds ($f_{\rm wind}=1.0$, though see above) and step overshooting of $\delta_{ov}=0.12$, 
to maintain consistency with the overshooting in models by \citet{Pols1998} and the \citet{Hurley2000} 
fits. At the end of the MS, this model has a mass of $M_{\rm TAMS}=94.0\msun$ with a He core mass of 
$M_{\rm core,TAMS}=66.6\msun$ (see Table~\ref{tab:MESA_mod}). Post-MS expansion leads to a maximum 
radius of $R_{\rm max}=1968\rsun$ at the end of our simulation, at which point the star is still 
expanding as a red supergiant. A {\tt MESA} stellar model with $M_{\rm ZAMS}=131\msun$, $Z=0.006$, and 
$V_{\rm i}=250\kms$ results in $M_{\rm MS}=86.6\msun$ and $M_{\rm core}=58.5\msun$ and expands 
beyond $2000\rsun$ (see Table~\ref{tab:MESA_mod}).

The $M_{\rm ZAMS}=144\msun$ ($Z=0.006$) non-rotating {\tt StarTrack} or {\tt COMPAS} model produces 
$M_{\rm TAMS}=77.4\msun$ and $M_{\rm core,TAMS}=34.1\msun$. We decrease the winds during MS, keeping 
the original \cite{Vink2001} prescriptions, but decreasing the LBV winds to $f_{\rm lbv}=0.48$ to get 
a model with $M_{\rm TAMS}=94.2\msun$ and $M_{\rm core,TAMS}=44.5\msun$. Next, we increase the core 
size by $f_{\rm core}=1.5$ to get the target values: $M_{\rm TAMS}=94.2\msun$ and 
$M_{\rm core,TAMS}=66.7\msun$ in the population synthesis codes. Note that such a massive star is 
already luminous ($L \sim 3\times10^6 L_{\odot}$) and cold enough ($T_{\rm eff} \sim 30,000$ K) to be 
beyond the Humphreys-Davidson limit~\citep{Humphreys1994} and subject to LBV winds on the MS. Applying 
the same calibration to $M_{\rm zams}=131\msun$ ($Z=0.006$) we obtain in population synthesis codes: 
$M_{\rm MS}=89.5\msun$ and $M_{\rm core}=62.3\msun$. We apply this calibration for all metallicities. 
However, one should note that some observations may be in contradiction of metallicity-independent LBV 
winds~\citep{Gilkis2021}.

\begin{table}[h]
\caption{
Properties of {\tt MESA} models used to calibrate the population synthesis simulations (the two top 
rows) as well as a few models with increased wind mass-loss or core overshooting (see 
Sec.~\ref{sec:postMSexpansion}). We show the maximum radius reached in our simulation as well as the 
total mass and the He core mass at the end of the MS phase.} 
\begin{tabular}{lccc}
\hline\hline
model   & $R_{\rm  Max}$ & $M_{\rm TAMS}$ & $M_{\rm  core,TAMS}$\\
        & [$\rsun$] & [$\msun$] & [$\msun$]\\
\hline\hline
144 $\msun$ ($\delta_{ov}=0.12$, $f_{\rm wind} = 1$) & $>$1968 & 94 & 67 \\
131 $\msun$ ($\delta_{ov}=0.12$, $f_{\rm wind} = 1$) & $>$2000 & 87 & 58 \\
\hline \hline
144 $\msun$ ($\delta_{ov}=0.33$, $f_{\rm wind} = 1$) & 46 & 62 & 61 \\
131 $\msun$ ($\delta_{ov}=0.33$, $f_{\rm wind} = 1$) & 41 & 59 & 58 \\
144 $\msun$ ($\delta_{ov}=0.12$, $f_{\rm wind} = 1.5$) & 40 & 57 & 55 \\
131 $\msun$ ($\delta_{ov}=0.12$, $f_{\rm wind} = 1.5$) & 40 & 63 & 53 \\
\hline
\hline
\end{tabular}
\label{tab:MESA_mod}
\end{table}

\subsubsection{Calibration of post-MS expansion}
\label{sec:postMSexpansion}

When considering the future fate of the Mk~34 system, a key question is whether its very massive 
components will expand after the end of the MS and initiate a mass-transfer interaction or whether they 
will lose their hydrogen envelopes already during the MS and smoothly transition to become compact 
helium stars, avoiding any RLOF. The two crucial aspects that affect the degree of the post-MS expansion 
of very massive stars is the amount of core overshooting and the strength of stellar winds. Here, we 
explore this by computing a small grid of {\tt MESA} models with different overshooting and wind 
assumptions.

For the calibration of TAMS properties of $131\msun$ and $144\msun$ models, we assumed a modest core 
overshooting of $\delta_{ov}=0.12$, following the calibration to low-mass stars in open clusters by 
\citet{Pols1998}. More recently, \citet{Choi2016} found the best agreement with the properties of 
the Sun for a {\tt MESA} model with $\delta_{ov}=0.16$. However, there is an increasing amount of 
evidence that core overshooting could be significantly larger in the case of massive stars 
\citep[e.g.,][]{Brott2011,Castro2014,Claret2018,Scott2021}. In particular, the calibration by 
\citet{Brott2011} to match the observed drop in rotational velocities of post-MS B stars
\citep[although see][]{Vink2010} resulted in $\delta_{ov}=0.33$, a value that has become widely used 
to compute stellar models of massive stars in the recent years. On the other hand, there is no 
observational calibration of core overshooting in the case of very massive stars of masses above 
$100\msun$. As such, we explore six different $\delta_{ov}$ values from a wide range between $0.12$ 
and $0.5$.

The Dutch wind scheme in {\tt MESA} incorporates the \citet{Vink2001} prescription for optically 
thin line-driven winds of hot MS stars. However, as noted above, very massive stars possess a 
sufficiently high luminosity to mass ratio on their MS that they approach their Eddington limit, 
leading to high mass-loss rates \citep{Grafener2008, Vink2011b, Bestenlehner2014, Bestenlehner2020a}. 
Here, we attempt to correct for this by simply increasing the wind scaling factor 
$f_{\rm wind}$ from $1.0$ to $1.5$ or to $2.0$. 

We compute a grid of $131\msun$ and $144\msun$ models with the above variations in overshooting and 
winds (and all the other assumptions same as in our calibration models in Sec.~\ref{sec:mesa_calib}). 
All the results are shown in Appendix~\ref{sec.app} and a few selected examples in 
Table~\ref{tab:MESA_mod}. In short, we find that any model with overshooting of $\delta_{ov}=0.33$ 
or higher, or a wind multiplication factor $f_{\rm wind} \geq 1.5$ evolves to become a helium WR 
star already by the end of MS, avoiding radial expansion beyond $100\rsun$ and any RLOF interaction 
in the Mk~34 binary. This result is at the basis of the quasi-single evolutionary scenario for Mk~34, 
see Sec.~\ref{sec.single}.

\subsubsection{Calibration of envelope binding energies} 
\label{sec:Ebindcalib}

Recent studies by \cite{Klencki2021} and \cite{Marchant2021} have shown that the envelope binding 
energies used in {\tt StarTrack} and {\tt COMPAS} (i.e. $\lambda$ scaling following \cite{Xu2010,
Dominik2012}) may be severely underestimated in the case of massive stars with outer radiative 
envelopes. Note that population synthesis codes do not perform CE evolution for Hertzsprung gap 
stars (radiative outer envelope) under standard assumptions on input physics, but typically the CE is 
applied for core-helium burning stars even for those with outer radiative envelopes. 

To explore the effect of revised binding energies on the future fate of Mk~34, whenever our
{\tt StarTrack} binary evolution calculation predicts a CE phase to occur, we use {\tt MESA} to 
compute a detailed stellar model of the donor star. We then follow the method outlined in 
\citet{Klencki2021} to integrate through the envelope of the {\tt MESA} model and compute its 
binding energy. When matching the properties of a {\tt MESA} model with those from {\tt StarTrack}, 
we ensure that the CE donor has the same helium core mass and the same radius but allow for a lower 
envelope mass in the {\tt MESA} model, so that we may be under- but never over-estimating the 
envelope binding energy. This allows for conservative statements on inability of a binary to eject the
donor's envelope and the CE survival.

\subsection{Pavlovskii calculations}
\label{sec.pavlovskii}

Models presented below are obtained with modified {\tt StarTrack}. In particular, we use more 
restrictive criteria for the CE development~\citep{Pavlovskii2017}, and allow more binaries to 
evolve through stable mass transfer instead~\citep{Olejak2021}. The new criteria are applied to 
{\em (i)} H-rich post MS donor stars, {\em (ii)} for initial masses larger than $18\msun$, 
{\em (iii)}  when the mass ratio (companion to donor mass at CE onset) fulfils the condition 
$q_{\rm CE}<q_{\rm crit}$ for CE to develop, where $q_{\rm crit}=0.19-0.36$ depending on donor mass 
and metallicity, and {\em (iv)} when the donor's radius at the onset of CE fulfils specific criteria 
(shown at Fig. 2 and 3 of \cite{Olejak2021}) for CE to develop. These new criteria lead to the 
emergence of BH-BH formation channels without CE in {\tt StarTrack} simulations~\citep{Olejak2021}. 
This is the same channel that was proposed by~\cite{Heuvel2017} and that is also found in 
{\tt COMPAS} simulations~\citep{Stevenson2019,Neijssel2019}. 

Models labeled "Pavlovskii1" and "Pavlovskii2" correspond to models "StarTrack1" and "StarTrack2" 
but with modified CE development criteria, respectively. 

In model "Pavlovskii3" we test different formulae for the loss of the angular momentum during mass 
transfer through the L2 point given by \cite{MacLoad&Loeb_2020}: 
$j_{\rm loss} =j_{\rm L2}=1.2^2 \frac{M_{\rm tot}^2}{M_{\rm don}M_{\rm acc}}$ instead of our 
standard $j_{\rm loss}=1.0$ \citep{Podsiadlowski1992}. We expect much higher loss of angular 
momentum with this modification, which may result in another potential fate for the future evolution 
of Mk~34 (see Sec.~\ref{sec.evol4}). The rest of input physics of "Pavlovskii3" corresponds to 
"Pavlovskii2". A change of the circularization scheme would only increase binary orbital separation.

\subsection{Quasi single star calculations}
\label{sec.single}

In this part we approximate the evolution of non-expanding components of Mk~34. This is driven by MESA 
models with large overshooting or increased stellar winds that do not show significant post-MS 
expansion (see Tab.~\ref{sec.app}). In particular, {\tt MESA} models with 
$\delta_{\rm ov}=0.33$ and standard winds ($f_{\rm wind}=1.0$) reach a maximum radius of $46\rsun$
for $M_{\rm ZAMS}=144\msun$ and $41\rsun$ for $M_{\rm ZAMS}=131\msun$. Additionally, important for 
the development of the PPSN/PSN, stellar models that do not expand may have very different TAMS helium core 
masses. For example, the $M_{\rm ZAMS}=144\msun$ model produces $M_{\rm core,TAMS}=61\msun$
for $\delta_{\rm ov}=0.33$ and $f_{\rm wind}=1.0$ or $M_{\rm core,TAMS}=30\msun$ for 
$\delta_{\rm ov}=0.4$ and $f_{\rm wind}=1.5$. The former model is possibly subject to a PPSN/PSN while 
the latter is not~\citep{Woosley2017,Farmer2020}.

In Sec.~\ref{sec.evol5} we explain our choice of models, showing how uncertainties can affect the 
future evolution of Mk~34.

\section{Examples of Mk~34 Future evolution} 
\label{sec.evol}

\begin{figure*}
\vspace*{0.4cm}
\hspace*{-0.4cm}
\includegraphics[width=1.0\textwidth]{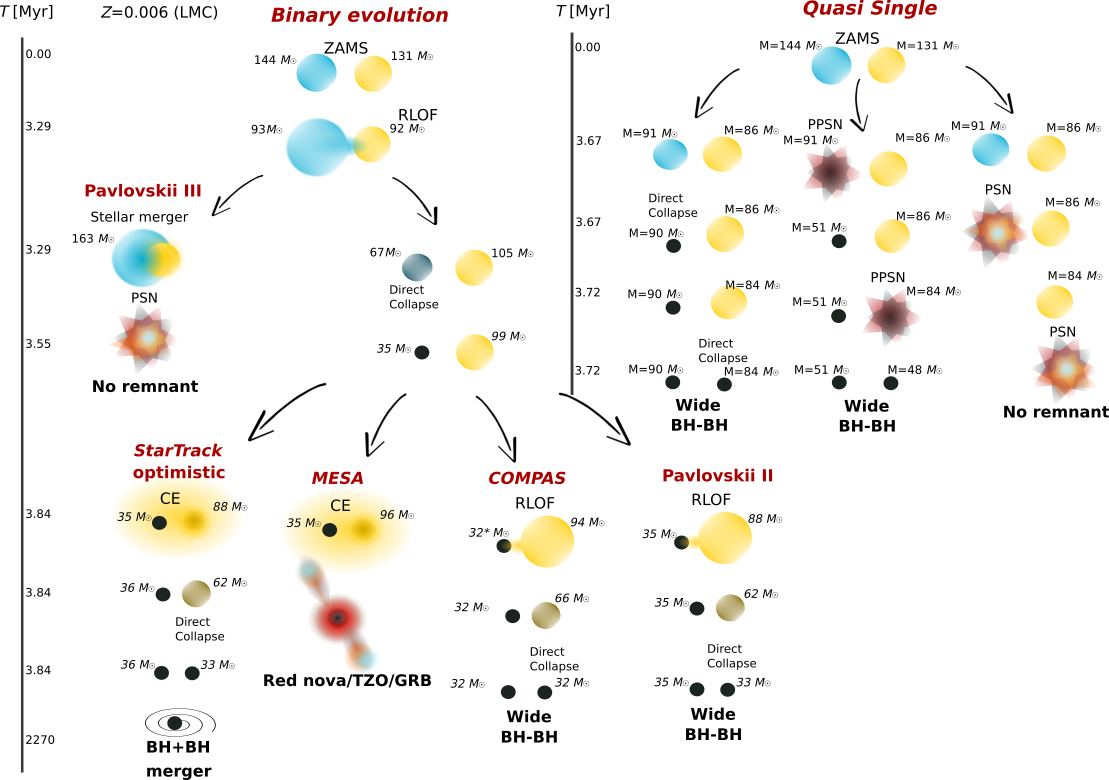}
\caption{
Future evolution of Melnick 34. The fate of this massive binary is subject to a number of stellar and 
binary evolution uncertainties. Depending on the adopted evolutionary model, Mk~34 may form a close or wide 
BH-BH system, a Thorne-\.Zytkow \citep[TZ --][]{Thorne1977} object or end its life in two pair-instability supernovae (PSN). 
RLOF: Roche lobe overflow, CE: common envelope, TTMT: thermal-timescale mass transfer, ZAMS: Zero 
Age Main Sequence, BH: black hole, GRB: gamma-ray burst, PPSN: pair-instability pulsation supernova. \\
$^\star$: {\tt COMPAS} track gives somewhat lower mass BH (formed out of an initially more massive 
star) than other binary scenarios. 
}
\label{fig.evol}
\end{figure*}

Various predicted models of Mk~34 future evolution are illustrated in Figure~\ref{fig.evol}, 
summarized in Table~\ref{tab.models}, and described below. 

\subsection{StarTrack models}
\label{sec.evol1}

A binary star resembling Mk~34 is evolved with the StarTrack2 model (Sec.~\ref{sec.startrack}). 
Star A (initially more massive) expands as it evolves and finally  periodically overfills its Roche lobe 
($R_{\rm a}=139\rsun$) at periastron passages. At this point the orbit expanded 
from the initial $a=760\rsun$ to $1125\rsun$ due to wind mass loss from both binary components 
($M_{\rm a}=94.2\msun$, $M_{\rm b}=91.9\msun$) while the eccentricity remained much unchanged 
($e=0.68$). The tidal circularization force is the strongest at periastron and we assume that 
the orbital motion is circularized to periastron distance and leads to normal stable RLOF on new 
($a=360\rsun$) circular orbit. RLOF leads first to TTMT which subsequently transforms to a NTMT. Star A is 
stripped almost entirely of its H-rich envelope ($M_{\rm a}=67.4\msun$, $M_{\rm a,core}=67.3\msun$) 
while star B accreted half of that lost envelope ($M_{\rm b}=105\msun$) while the rest of the mass has been lost 
from the binary. In response the orbit increased in size ($a=439\rsun$). Star A is a Wolf-Rayet 
with heavy wind mass--loss and at the end of its nuclear evolution its mass decreases to 
$M_{\rm a}=35.6\msun$ ($a=1370\rsun$, $M_{\rm b}=99.1\msun$). Star A collapses directly to a BH 
with a mass of $M_{\rm a}=35.2\msun$ ($1\%$ neutrino mass loss, no baryonic mass loss, no natal 
kick). Then star B evolves and expands to fill its Roche lobe in a circular orbit with 
$a=1515\rsun$ ($M_{\rm b}=88.5\msun$, $R_{\rm b}=700\rsun$, $T_{\rm b,eff}=8690$ K). This time 
in our standard approach, due to relatively high mass ratio ($q=88.5/35.2=2.5$) RLOF is evaluated 
to lead to a CE phase. We estimate the binding energy of star's A envelope ($\lambda=0.103$) to be low 
enough to be ejected at the cost of orbital energy. After envelope ejection star B becomes a massive 
stripped He core ($M_{\rm b}=62.0\msun$) and the orbit decays to $a=34.2\rsun$. During CE the first-formed 
BH accretes $\sim 0.5\msun$ ($M_{\rm a}=35.7\msun$). Star B, after a Wolf-Rayet wind mass loss 
($M_{\rm b}=33.7\msun$, $a=47.8\rsun$) collapses directly to a BH ($M_{\rm b}=33.3\msun$). After 
$t_{\rm evol}=4.1$Myr of binary evolution a close BH-BH binary is formed with a coalescence time of 
$t_{\rm coal}=10.0$Gyr.   

There are caveats in this scenario. Star B at the time of the RLOF onset has just finished core H-fusion and 
is a Hertzsprung-gap star in the transition to become a core-He burning giant. It was argued that 
such stars do not have a clear core-envelope structure and that the CE phase should always lead to merging of 
the donor star with its companion~\citep{Belczynski2007}. This finds some support in observations as the
predicted BH-BH merger rates that allow for such a scenario as presented above are too high to match the
empirical LIGO/Virgo estimate (see submodels A in Table 4 of~\cite{Belczynski2020b}). In addition, 
star B has an outer radiative envelope: at the time of RLOF and with surface properties 
$R_{\rm b}=700\rsun$, $T_{\rm b,eff}=8690$K $\rightarrow L=2.5\times 10^6 {\rm ~L}_\odot$, it is 
well above the effective temperature threshold below which stars at LMC metallicity have convective 
envelopes ($T<3900$K; see Fig.6 of~\cite{Klencki2020}). \citet{Klencki2021} argued that massive 
radiative-envelope giants have binding energies that are too high to allow for a successful CE 
ejection in BH-BH merger progenitor binaries. In {\tt StarTrack} models, the estimate of the 
binding energy of star B with  the $\lambda$ formalism is only an approximation that is needed for the use 
in rapid population synthesis models. It should be also stressed that in the standard input physics of {\tt StarTrack} 
models we would allow no Hertzsprung-gap star to survive a CE phase no matter what is our estimate of the
star binding energy ($\lambda$) or assumed efficiency of orbital energy transfer to the envelope
($\alpha$). This is why we call the {\tt StarTrack} models used here as optimistic scenarios. We test 
this optimistic estimate with MESA in Sec.~\ref{sec.evol3}. 

In StarTrack1 model a similar scenario develops. However, since the stars and their cores are less 
massive the BH masses are $\sim 10\msun$ smaller than in model StarTrack2 (see Tab.
~\ref{tab.models}).   

In StarTrack3 model we adjust evolutionary parameters in such a way that CE survival is less 
caveated than in model StarTrack2. We alter circularization process and we change RLOF 
parameters setting mass transfer/loss to obtain wider binary than in StarTrack2 model. This 
allows star B to expand more before it initiates CE phase: $R_{\rm b}=1439\rsun$, $a=3141\rsun$, 
$M_{\rm b}=85.9\msun$, $M_{\rm b,core}=60.1\msun$, $T_{\rm b,eff}=5986$ K, $\lambda=0.050$. This 
star has almost convective envelope, but not quite so. If we perform CE energy balance with the 
above parameters this system survives CE and forms close BH-BH binary at $t_{\rm evol}=3.9$ 
Myr and with $t_{\rm coal}=11.7$ Gyr. Yet, the same caveats remain as for StarTrack2 model.

\subsection{COMPAS models}
\label{sec.evol2}

With the initial masses, orbital period and eccentricity of Mk~34 (see Sec.~\ref{sec.calc}) 
we evolve COMPAS1, COMPAS2 and COMPAS3 models with the individual variations specified in 
Sec.~\ref{sec.compas}. As in Sec.~\ref{sec.evol1}, we will always refer to the originally more 
massive star (with $M_{\rm ZAMS}=144\msun$) as star A. 

In COMPAS2 model, star A ends core-H burning with a total mass of $94.2\msun$, while the stellar 
winds increase the separation to $a=1165.3\rsun$.  As star A leaves the MS, its core mass is 
calculated to be $M_{\rm a,core}=66.8\msun$. Star A overfills its Roche lobe soon during post-MS 
evolution losing its envelope ($27.4\msun$) in a stable (fully conservative) TTMT RLOF phase. 
This is the outcome of the binary being fairly wide and the donor being an evolved (post-MS) star, 
as commented on in Sec.~\ref{sec.compas} (see also~\cite{Schneider2015}). Eccentricity remains 
unchanged ($e=0.68$) during this phase and the orbit expands to $a=1374.9\rsun$. Star B, which had a 
total mass of $91.8\msun$ right before the mass transfer, becomes a $119.2\msun$ MS star. Being 
stripped off its H-rich envelope, star A enters the naked helium-star (WR) phase with mass
$M_{\rm a}=66.8\msun$. Evolution continues while both stars are losing mass in winds. Star A with
mass $M_{\rm a}=35.4\msun$ undergoes direct core-collapse, forming a BH of mass 
$M_{\rm a}=31.8\msun$, while star B with $M_{\rm b}=115.3\msun$ is still on the MS. The orbital 
separation, right after the formation of BH A, becomes $a=1712.1\rsun$. Star B leaves the MS with mass 
$M_{\rm b}=93.7\msun$ and core mass $M_{\rm b,core}=66.3\msun$ when the orbital separation is 
$a=2006.6\rsun$. Shortly thereafter star B fills its Roche lobe, loses most of its H-rich  envelope 
($\sim 27.3\msun$) and becomes a naked helium star in stable TTMT RLOF. The post-RLOF orbital 
separation is decreased to $a=918.1\rsun$. Eddington limited accretion allows BH A to gain only 
$\sim 3.7\times10^{-5}\msun$. The binary at the onset of this RLOF, though has slightly smaller 
separation than the previous RLOF phase, is still fairly wide and results in a stable TTMT RLOF
despite rather high mass ratio ($q=93.7/31.8=2.9$). It is noted that both RLOF phases noted in 
COMPAS2 model are fairly similar to Pavlovskii2 model (see Sec.~\ref{sec.evol4}).  However, the 
binary remains eccentric through both RLOF phases ($e=0.673$). Star B, with a mass 
$M_{\rm b}=35.2\msun$, undergoes direct core-collapse and forms a BH of $31.7\msun$. COMPAS2 model 
creates a BH-BH system, which at the second BH formation has a separation $a=1372.6\rsun$ and an 
eccentricity of $e=0.658$. This wide BH-BH system does not merge in a Hubble time. 

The evolution within COMPAS1 model is fairly analogous  to COMPAS2 model. However, higher wind mass-loss 
and less internal mixing leads to formation of a much less massive BH-BH system. BHs in this 
model are $\sim 20\msun$ while in COMPAS2 they are $\sim 32\msun$. The orbital separation at BH-BH 
formation is $a=740.6\rsun$ while the eccentricity remains virtually unchanged ($e=0.68$) resulting in a
coalescence time longer than the Hubble time. 

Results of evolution in the COMPAS3 model are also broadly similar to those of the COMPAS2 model (see 
Tab.~\ref{tab.models}). The additional condition of pre-RLOF orbital circularization, however,
changes a few key points. The first RLOF (stable, TTMT, star A to star B) decreases the orbital 
separation from $11165.6\rsun$ to $739.2\rsun$ and the eccentricity becomes $e=0.0$. As star B 
evolves, stellar winds increase this orbital separation to $a=1097.3\rsun$ right before the second 
RLOF (stable, TTMT, star B to BH A). At this point, the eccentricity is $e=0.023$, increased from 
the previous circularization at the formation of the first BH. The second RLOF decreases the separation 
to $a=501.9\rsun$, and again the orbit is circularized. The binary orbital separation at the time of 
formation of the BH-BH system (with $\sim 32\msun$ BHs) becomes $a=779.6\rsun$ while the orbital 
eccentricity is negligible ($e=0.054$). Though COMPAS3 model evolution  decreases the 
binary orbital separation substantially, we note that this change is not significant enough to create a close 
double BH system that merges in a Hubble time. 

We note that COMPAS3 creates the closest BH-BH binary obtained in the three 
COMPAS models. We note that to create a similarly circularized double BH system with the same masses, that 
merges within a Hubble time, the orbital separation at BH-BH formation can at most be about 
$46.7\rsun$~\citep{Peters1964}. A highly eccentric orbit can also decrease the merger time. However, 
for a BH-BH system with same masses and orbital separation as in the COMPAS3 model, this cut-off 
eccentricity should be at least $e=0.98$. 

Interestingly the structure of the BH X-ray binary Cyg X-1 (though a less massive system than Mk42) was 
used in StarTrack~\citep{Wiktorowicz2014} and COMPAS~\citep{Neijssel2021} calculations to argue that the 
future evolution of Cyg X-1~\citep{MillerJones2021} may also lead to a wide BH-BH system which will not
merge in a Hubble time.

\subsection{MESA models}
\label{sec.evol3}

In this section we use the {\tt MESA} code to check the outcome of the CE phase encountered in
the three {\tt StarTrack} models from Sec.~\ref{sec.evol1}
\citep[see Sec.~\ref{sec:Ebindcalib} and][for the method]{Klencki2021}.

In the model MESA2 we evolve a star with $M_{\rm zams}=133\msun$, overshooting $\sigma_{\rm ov}=0.2$ and 
Dutch winds with $f_{\rm wind}=1.5$. This model produces at some point of its post-MS evolution a 
star with a mass of $M=80.9\msun$ and a He-core mass of $M_{\rm core}=62.2\msun$, radius of $R=700\rsun$ 
$T_{\rm eff}\approx 9000$K (an outer radiative envelope; the envelope would not become convective until 
least $T_{\rm eff}\lesssim4500$K \cite{Klencki2020}). 
At this point we calculate the envelope binding energy (obtained from integration of the mass distribution over the
entire envelope) corrected for the internal energy of the envelope, then we subtract the BH accretion 
luminosity that effectively lowers binding energy,  to obtain $E_{\rm bind}=6.67\times 10^{50}\erg$. 
This translates to $\lambda_{\rm MESA}=0.012$. This model resembles the star B at the onset of CE in the
StarTrack2 example of evolution. The orbital energy at the onset of CE is 
$E_{\rm orb,i}= - 0.04 \times 10^{50}\erg$ and the post-CE separation is $a=6.2\rsun$ (corresponding to a
post-CE $E_{\rm orb,f}= - 6.81 \times 10^{50}\erg$). This was obtained under the assumption of a $100\%$ 
efficiency of the orbital energy transfer to unbind the envelope ($\alpha=1.0$). The radius of the exposed 
core of star B is $2.84\rsun$ (~\cite{Hurley2000} formulae) while its new Roche lobe is only $2.64\rsun$ 
and we assume a CE merger in such case. In the StarTrack2 model the binding energy ($\lambda=0.103$) was 
underestimated as it scales with $\propto 1/\lambda$ by factor of $\sim 9$ as compared with detailed 
MESA estimate ($\lambda=0.012$).  

Note that since the BH accretes part of the envelope during the CE inspiral (for the numerical treatment of this 
process see Appendix in~\cite{Belczynski2002}), the binary does not need to balance the entire binding 
energy of the envelope ($E_{\rm bind}$) with the orbital energy ($E_{\rm orb,f}-E_{\rm orb,i}$). 
However, this has no influence on our conclusion above. Accretion onto the BH is estimated at the level 
of $\sim 0.5\msun$ (see Sec.~\ref{sec.startrack}), while this MESA model underestimates the mass of the
stellar envelope found in StarTrack2 simulation by $\sim 7.4\msun$ ($M_{\rm env}=26.1\msun$ in 
StarTrack2 simulation, and $M_{\rm env}=18.7\msun$ in the above MESA simulation).

The other two population synthesis models from StarTrack also produce CE mergers (Thorne-\.Zytkow 
objects) if MESA binding energy estimate is used. In StarTrack1 model the CE donor was estimated to 
have $\lambda=0.050$, while the MESA calculation gives $\lambda=0.007$. In StarTrack3 model 
$\lambda=0.050$ as contrasted with MESA estimate of $\lambda=0.008$.

\subsection{Pavlovskii models}
\label{sec.evol4}

A binary star resembling Mk~34 is evolved with the Pavlovskii2 model (Sec.~\ref{sec.pavlovskii}). 
Evolution to RLOF initiated by star B is the same as in the StarTrack2 model ($\lesssim 1\%$ 
differences in binary parameters are numerical). However, here the binary undergoes stable a RLOF: 
TTMT instead of CE. During mass exchange mass/loss the orbital separation changes from 
$a=1507\rightarrow928\rsun$ and star B is stripped from its H-rich envelope 
($M_{\rm b}=87.7\rightarrow62.3\msun$) becoming a massive Wolf-Rayet star. Accretion onto the BH is 
negligible as the mass transfer was highly supper-Eddington ($M_{\rm a}=35.2\rightarrow35.2\msun$) 
and the loss of angular momentum (given the mass ratio at the onset of RLOF: $q=87.7/35.2=2.5$) 
causes orbit to decrease in size by a factor of $1.6$; this may be compared with the orbital decrease 
by a factor of $63$ during CE in the StarTrack2 model. After Wolf-Rayet wind mass loss 
($M_{\rm b}=33.9\msun$, $a=1427\rsun$) star B collapses directly to a BH ($M_{\rm b}=33.3\msun$). 
After $t_{\rm evol}=4.16$Myr of binary evolution a wide BH-BH binary is formed with a coalescence 
time of $t_{\rm coal}=6.0\times 10^6$Gyr.   
 
In the Pavlovskii1 model, the binary follows a similar evolution, but a lower-mass wide BH-BH binary forms 
($21.9+22.1\msun$) due to stronger winds and lower core masses adopted in this model (see 
Tab.~\ref{tab.models}). 

In Pavlovskii3 model, due to strongly increased loss of the orbital angular momentum during TTMT, 
the system does not survive the first TTMT. It ends with a stellar merger of the Hertzsprung gap star 
donor ($M_{\rm a}=80.2\msun$) with its main-sequence star companion ($M_{\rm b}=98.9\msun$). The 
actual mass and the fate of the stellar-merger product is uncertain. Both observations and 
simulations of stellar mergers are usually related to low-mass stars, which are not BH progenitors 
\citep{Lombardi_Jr__2002,2016A&A...592A.134T}, or are calculated for dynamical collisions in 
dense stellar clusters~\citep{Glebbeek_2013}. It seems that a rather low mass fraction is lost 
during stellar mergers~\citep{Lombardi_Jr__2002,2006ApJ...640..441L,Glebbeek_2013}. Assuming that 
the merger product in our simulation will become a Hertzsprung-gap star with the mass of 
$\sim 163\msun$ (similarly to the scheme used in \cite{Olejak2020a} with $20\%$ of the less massive 
star being ejected during the merger) the single star will end its evolution either as {\em (i)} PSN 
leaving no remnant if classical PSN models are used~\citep{Woosley2017,Leung2019}, or {\em (ii)}
as a single $\sim 30-40\msun$ BH, if non-standard PSN models are used
~\citep[see Fig.1 of][and references therein]{Belczynski2020c}. 

The amount of angular-momentum loss through the L2 Lagrangian point adopted in the Pavlovskii3 model,
$j_{\rm loss}=j_{\rm L2}=1.2^2\frac{M_{\rm tot}^2}{M_{\rm don}M_{\rm acc}} \in 
[5.76,5.82]$,\footnote{This range corresponds to changing mass of donor and accretor during RLOF.} is 
considered to be an upper limit whereas the standard {\tt StarTrack} $j_{\rm loss}=1.0$ used in 
models Pavlovskii1 and Pavlovskii2 is instead close to the lower limit as indicated by 
\cite{MacLeod_2018} and \cite{MacLoad&Loeb_2020}. The maximal possible $j_{\rm loss}$ which allows 
to avoid a stellar merger during the first TTMT and would lead to the formation of a wide BH-BH 
binary from Mk~34 is $j_{\rm loss} \in [3.74,3.92]$ ($65\%$ of $j_{\rm L2}$ of 
\cite{MacLoad&Loeb_2020}). This demonstrates that even with increased angular momentum losses, it 
is possible to form either a wide BH-BH binary ($t_{\rm delay}>t_{\rm hub}$) with minimal separation 
of about $a=609\rsun$ ($t_{\rm delay}=2.6\times 10^{5}$ Gyr $>t_{\rm hub}$) or a stellar merger
but not a close BH-BH system.

\subsection{Quasi single star evolution models}
\label{sec.evol5}

Based on {\tt MESA} models (see Sec.~\ref{sec.app}), current literature and simple estimates, we 
follow the future evolution of the Mk~34 binary with non-expanding stars. We put the two stars on an 
eccentric ($e=0.68$) and wide orbit ($a=780\rsun$). These stars lose $\gtrsim 100\msun$ during their 
MS life in stellar winds expanding the orbital separation ($a>1000\rsun$; see Sec.~\ref{sec.app}). 
At the post-MS closest encounter of these two stars (periastron), Roche lobe radii of both components 
are $R_{\rm lobe}>100\rsun$. The radii of both stars are $R<100\rsun$ for many {\tt MESA} models. There 
is no mass exchange between the stars.  

Depending on {\em (i)} the mass and core mass of non-expanding stellar models and {\em (ii)} the 
mass (in reality central temperature and density) and the range allowed for the onset of a PPSN/PSN, 
we can envision several different outcomes of Mk~34's future evolution. 

If both stars have core masses as high as $61\msun$ and $58\msun$ at TAMS (see {\tt MESA} models 
with $\delta_{ov}=0.33$ and $f_{\rm wind}=1.0$ in Sec.~\ref{sec.app}), these cores will reach 
$65\msun$ at the time of oxygen burning, which will then become explosive leading to PSN
~\citep{Woosley2017}. Each star gets disrupted, leaving no compact object remnant but producing 
luminous PSN supernova (model: QuasiSingle1; see Tab.~\ref{tab.models}).  

There is a significant caveat to the above prediction. According to recent studies
~\citep{Woosley2017,Limongi2018,Farmer2020,Costa2021,Farrell2021} very low metallicity stars can 
produce BHs with mass as high as $\sim 80-90\msun$. But there are also detailed {\tt MESA} stellar 
evolutionary models that allow for the formation of BHs with $\sim 70\msun$ avoiding a PPSN/PSN
~\citep{Belczynski2020a} at high metallicity. If this scenario is adopted then it is expected that  
$60+60\msun$ wide BH-BH binary would form ((model: QuasiSingle2).

For lower mass stars/cores at TAMS, we expect avoiding a PPSN/PSN and we predict the formation of a 
wide BH-BH binary. For example, if we take {\tt MESA} models with $\delta_{ov}=0.4$ and 
$f_{\rm wind}=1.5$ they will produce stars with $M=32\msun$ ($M_{\rm core,TAMS}=30\msun$) and $M=32\msun$ 
($M_{\rm core,TAMS}=30\msun$). These stars are not subject to PPSN/PSN and depending on the post-MS 
stellar wind mass loss will form $\sim 30\msun$ BHs (model: QuasiSingle3). The formation of a wide 
BH-BH binary, with coalescence time exceeding the Hubble time, is predicted. 

If we push {\tt MESA} models even further to higher overshooting and stronger winds 
($\delta_{ov}=0.5$ and $f_{\rm wind}=2.0$) we produce stars with $M=21\msun$
($M_{\rm core,TAMS}=19\msun$) and $M=21\msun$ ($M_{\rm core,TAMS}=19\msun$) at TAMS. This will also 
lead to the formation of a wide BH-BH binary but with $\sim 20\msun$ BHs at most (model: 
QuasiSingle4).

\section{Discussion}
\label{sec:dis}

We have investigated the future evolutionary tracks and fate of the most massive known binary system Mk~34. 
Several interesting possibilities seem to exist (see Tab.~\ref{tab.models}). However, it is
impossible to decide with certainty (due to various stellar and binary physics uncertainties) which 
predicted fate is the correct one (if any). 

If very massive stars at LMC metallicity and with moderate rotation expand during their post-MS 
evolution (expected for low overshooting) then we predict the following evolution sequence for 
Mk~34: 
\begin{equation}
RLOF_{\rm A}\ \rightarrow BH_{\rm A}\ \rightarrow  RLOF_{\rm B}\ \rightarrow TZ_{\rm A+B}/BH_{\rm B}
\end{equation} 
where indices ``A" and ``B" mark the more- and less-massive component of Mk~34 respectively, $BH_{\rm A/B}$ 
denotes the BH formation from a given component, and $TZ_{\rm A+B}$ means the formation of a Thorne-\.Zytkow 
object from both binary components in the second RLOF. The first RLOF (initiated by star A) is always 
found to be stable (TTMT/NTMT), while the second RLOF (donor: star B) can be either stable or 
dynamically unstable (CE). Additionally, the binary system may not survive the first RLOF while both stars 
merge forming a single star that will be either subject to PSN (no remnant) or will form a single BH: 
\begin{equation}
RLOF_{\rm A}\ \rightarrow single\ star\ \rightarrow PSN/BH
\end{equation}

If such massive stars do not expand (for example, because of significant overshooting) the future 
evolutionary history proceeds without any binary interaction: 
\begin{equation}
BH_{\rm A}/PPSN_{\rm A}/PSN_{\rm A}\ \rightarrow BH_{\rm B}/PPSN_{\rm B}/PSN_{\rm B} 
\end{equation}

Under very optimistic conditions (development and survival of CE initiated by a massive star with 
a radiative envelope and with comparable mass companion) Mk~34 may form a heavy BH-BH merger that 
would be a source of gravitational-waves. Depending on our assumptions on mass loss and mixing in 
stellar interiors we find formation of a $\sim 20+20\msun$ close BH-BH system that resembles 
LIGO/Virgo detection of 
GW190408\_181802 ($24.6^{+5.1}_{-3.4}+18.4^{+3.3}_{-3.6}\msun$:~\cite{Abbott2021a}) or 
$\sim 30+30\msun$ BH-BH system that would look similar to 
GW150914 ($35.6^{+4.8}_{-3.0}+30.6^{+3.0}_{-4.4}\msun$:~\cite{DiscoveryPaper}) or to 
GW190828\_063405 ($32.1^{+5.8}_{-4.0}+26.2^{+4.6}_{-4.8}\msun$:~\cite{Abbott2021a}).
If this does not work, the formation of such LIGO/Virgo BH-BH mergers can still be obtained with more 
realistic CE input physics in the isolated binary evolution~\citep{Belczynski2016b,Spera2019,Patton2021}. 

The detailed evolutionary estimates of a very massive star envelope binding energy do not allow for 
a CE survival of the Mk~34 descendant binary even under very conservative assumptions ($100\%$ 
orbital energy used to eject the envelope with the help of internal energy of the gas and accretion 
luminosity from the inspiralling BH). If this is taken into account, then instead of forming a close 
BH-BH binary in CE scenario, we encounter the formation of a Thorne-\.Zytkow object (BH sinks 
into the center of the post-MS massive star). This single object would first appear as a post-MS 
massive star (most likely a classical Wolf-Rayet star), that starts to expand, cooling off and getting redder 
as the envelope puffs up in response to the BH inspiralling in a H-rich envelope. Once the BH sinks into the star's  
core (the majority of star mass at this point) the accretion of helium is extremely rapid 
\cite[$\gtrsim 1\msun \ {\rm min}^{-1}$, as in a collapsar engine;][]{Fryer1998} and the core disappears 
and then the rest of the star is accreted as well and the object disappears entirely from sky. Such a
transient should be visible in optical/infrared (initial expansion of the envelope before it collapses 
onto the BH), although at this moment there are no available calculations of the light curve or spectra for 
such heavy mergers ($\sim 100\msun$ post-MS star and $\sim 20-30\msun$ BH). Observationally, various 
red novae/transients were proposed to be the outcome of CE mergers~\citep{Tylenda2005,Ivanova2013b,
Kaminski2015,MacLeod2017b}. It may be even possible that such a merger would lead to a gamma-ray burst 
(GRB). If there is enough angular momentum in the He core and BH then such a configuration may lead 
to formation of jets powering a GRB~\citep{Zhang2001}. The angular momentum transport in stellar 
interiors of massive stars is not fully constrained, although low effective spins of LIGO/Virgo 
BH-BH mergers seem to indicate efficient angular momentum transport in massive stars
~\citep{Spruit2002,Fuller2019a,Bavera2020,Belczynski2020b}. In such a case a slowly spinning He 
core--BH system would have only a small chance of producing a GRB. Another obstacle in producing 
a GRB in this case is $\sim 30\msun$ of H-rich envelope for jets to punch through. Yet, there 
are signs that such jets do form in massive stars in CE mergers and they try to breakout from 
stellar interiors~\citep{Thone2015}.

Another possibility for the future evolution of Mk~34 is to avoid the CE phase entirely, even if both 
stars in this binary expand. Since the orbital separation is not very large for this system, any 
RLOF encountered in the evolution is bound to happen when the donor star is not too large (a radiative 
envelope). Additionally, since the stars in Mk~34 are of similar mass, any RLOF is not bound to happen 
at extreme mass ratio. Taking this into account, this binary may evolve through two episodes of 
stable RLOF (first initiated by initially more massive star, and then by the other star). Although 
such RLOF episodes may decrease the orbital separation, such orbit shrinking will be not large enough 
to lead to the formation of a close BH-BH binary (i.e., with a merger time smaller than the Hubble time)
for the initial binary configuration of Mk~34. Instead, a wide BH-BH binary forms with separation 
as large as $\gtrsim 1000\rsun$. Such a descendant of Mk~34 cannot be a LIGO/Virgo source, but could 
possibly be detected by microlensing observations. The magnification of the source in such a microlensing 
event would last months and would include bumps typical of a binary lens and would be potentially 
detectable in LMC. 

Finally, it is also possible that neither of the stars in Mk~34 will experience any significant expansion in 
their post-MS evolution. In such a case, Mk~34 expands due to wind mass loss from both stars and forms 
a wide system when both components end their nuclear evolution (separation of $\sim 1000\rsun$). 
Depending on the highly uncertain mixing physics and not fully constrained nuclear reaction rates 
(overshooting, rotation, convection, carbon-fusion) that set temperature/density in stellar cores,
massive components of Mk~34 may or may not be a subject to significant mass loss associated with 
PPSN during oxygen burning. We predict the formation of a wide BH-BH system with comparable-mass BHs in 
the mass range $\sim 40-90\msun$. Alternatively, both stars in Mk~34 may be subject to a full-fledged 
pair-instability and get disrupted in luminous PSNe's~\citep{Higgins2021}.  

To summarize, we conclude that we cannot yet predict the fate of a massive binary such as Mk~34. The involved 
stellar and binary physics uncertainties are still too overwhelming. However, our study offers several 
conditional statements that shed light on the future evolution of this massive binary. If the stars in 
Mk~34 expand in their post-MS evolution then they are bound to initiate two RLOF interactions (one 
by each component). Although the first interaction is always stable and does not threaten the survival of this 
system, the second one is more problematic. It is not at all clear if the second interaction 
will be dynamically stable. If it is not, then a CE phase develops and the most likely fate of the system is then the 
merger of the two binary components, possibly associated with a red nova or a GRB. If the second 
interaction is stable then the RLOF will lead to the formation of a wide and massive BH-BH system with a
merger time much larger than the Hubble time. Such a system is not a potential LIGO/Virgo source of 
high-frequency gravitational waves, but it may produce a microlensing event. If the stars in Mk~34 do not 
expand, which is also allowed by the current detailed evolutionary models, then we predict either the
formation of a wide and potentially very massive BH-BH system, or the spectacular death of both stars in 
luminous pair-instability supernovae that leave no BHs behind.

\section{Conclusion}
\label{sec.con}

For any given origin scenario {\em (i)} there is a large number of input physics uncertainties or 
even unknowns (model parameters), {\em (ii)} the implementation of the physical processes that 
involve these uncertainties in the numerical codes are far from being based on first-principle 
physics, so even probing the full range of a given parameter might not get the right answer, and 
{\em (iii)} there are more parameters and thus uncertainties than are commonly realized (for
example, there are at least $\sim 30$ parameters in the isolated binary evolution, even though it is 
usually thought that a small subset of them are the most important in determining binary outcomes). 
Thus strong conclusions are unjustified at this time.

As an example of how biases could enter the model comparisons, suppose that we use the rapid 
supernova engine model of \citet{Fryer2012}, which naturally produces a $\sim 2-5\msun$ mass gap 
between neutron stars and black holes.  Then the discovery of the $2.6\msun$ object in GW190814 
would rule strongly against isolated binary evolution and in favor of another channel, such as 
primordial black holes or multiple-generation mergers in dense stellar systems (e.g., two neutron 
stars could merge to make a $\sim 2.6\msun$ black hole).  But perhaps the delayed supernova engine 
model of \citet{Fryer2012} is a better description; in this model the compact objects with mass 
$2.6\msun$ are produced naturally and the isolated binary formation channel is perfectly viable. Or 
perhaps some other supernova model is selected by Nature, which would change the Bayes factor 
between the models that are considered.  Similar considerations apply to the high-mass merger 
GW190521, which is consistent with binary stellar evolution given the substantial uncertainties 
\citep{Farmer2020,Belczynski2020c,Kinugawa2020,Vink2021a,Costa2021,Mehta2021}. 

We have focused on the specific system Mk~34, but our caveat extends to analyses of the full 
population. For example, \cite{Olejak2021} explored the effects on the BH-BH population that stem 
from different treatments of the common envelopes. Over the range of models they studied, the 
population characteristics varied drastically. For example, the BH-BH merger rate varied from 
$18\gpy$ (consistent with the current LVC estimate of $15.3-38.8\gpy$:~\cite{LIGO2020a}) to 
$88\gpy$. The BH mass distribution can be consistent with the LVC estimate ($\propto M^{-1.5}$ 
below $\sim 40-50\msun$ and $\propto M^{-5.3}$ at higher masses), or very inconsistent with the 
estimate ($\propto M^{+2.7}$ for $M<15\msun$ and $\propto M^{-3.3}$ for heavier black holes). The 
mass ratio distribution can have one peak or two peaks. To reiterate, even this broad range of 
predicted population characteristics does not include many other possible variations of aspects of 
stellar and binary evolution.

What must be done to reach a stage in which we \textit{can} draw firm conclusions about the BH-BH 
system origins? More and better data will obviously help: for example, if multiple events point to a 
compact object in the $\sim 2-5\msun$ range then this tells us that the lower mass gap is not a 
major feature of the mass distribution. Rare individual events, if they have definitively established 
properties, can point to particular origins. For example, an event with many cycles that is clearly 
highly eccentric would favor a dynamical origin, and a compact object with a mass $<0.5\msun$ would 
signify a primordial black hole. But we emphasize that a detailed work on the physics of BH 
formation in each case is essential: statistical analyses must be grounded in both thorough and 
accurate physics and astrophysics.

\vspace*{-0.2cm} \acknowledgements

We thank Lukasz Wyrzykowski, Tomasz Kami{\'n}ski, Chris Fryer and Daniel Holz for useful comments on 
the manuscript. KB, AR, AO acknowledge support from the Polish National Science Center grant Maestro 
(2018/30/A/ST9/00050).  DC and SS acknowledge the support of the Australian Research Council Centre 
of Excellence for Gravitational Wave Discovery (OzGrav), through project number CE170100004. This 
work made  use  of  the  OzSTAR  high  performance  computer which is funded by Swinburne University 
of Technology and the National Collaborative Research Infrastructure Strategy (NCRIS). MCM 
acknowledges support from NASA ADAP grant 80NSSC21K0649.  He performed part of his work on this 
paper at the Aspen Center for Physics, which is supported by National Science Foundation grant 
PHY-1607611. JPL was supported in part by a grant from the French Space Agency CNES.

\bibliography{biblio}

\begin{appendices}
\section{Winds-overshooting grid}
\label{sec.app}

\begin{table}[!htb]
\caption{Maximum radius and stellar mass, core mass and envelope mass at TAMS from different 
combinations of overshoot fractions and Dutch winds scale factors for the $144 \msun$ ZAMS 
star (left) and the $131 \msun$ ZAMS star (right).}
\begin{tabular}{|lcccc|}
\hline\hline
model         & $R_{Max}$ & $M_{TAMS}$ & $M_{core,TAMS}$ & $M_{env,TAMS}$ \\
              & [$\rsun$] & [$\msun$] & [$\msun$] & [$\msun$] \\
\hline\hline
&  &  &  & \\
$\delta_{ov}$ 0.12 &  &  &  & \\
$f_{\rm wind}$  1    & $>$1967.640 & 94.018 & 66.578 & 27.440  \\    
$f_{\rm wind}$  1.5  & 39.694 & 56.672 & 55.423 & 1.249  \\    
$f_{\rm wind}$  2    & 25.870 & 30.662 & 28.014 & 2.648  \\ 
&  &  &  & \\        
$\delta_{ov}$ 0.16 &  &  &  & \\
$f_{\rm wind}$  1    & $>$2000 & 93.139 & 68.443 & 24.696  \\    
$f_{\rm wind}$  1.5  & 34.663 & 47.202 & 45.412 & 1.790  \\    
$f_{\rm wind}$  2    & 24.192 & 27.978 & 25.227 & 2.750  \\ 
&  &  &  & \\        
$\delta_{ov}$ 0.2 &  &  &  & \\
$f_{\rm wind}$  1    & $>$1129.641 & 91.070 & 70.021 & 21.049 \\    
$f_{\rm wind}$  1.5  & 31.430 & 41.892 & 39.779 & 2.113  \\    
$f_{\rm wind}$  2    & 23.346 & 26.517 & 23.794 & 2.723  \\ 
&  &  &  & \\        
$\delta_{ov}$ 0.33 &  &  &  & \\
$f_{\rm wind}$  1    & 46.046 & 62.221 & 60.782 & 1.438  \\    
$f_{\rm wind}$  1.5  & 26.474 & 34.516 & 32.304 & 2.212  \\    
$f_{\rm wind}$  2    & 21.062 & 23.249 & 20.738 & 2.511  \\ 
&  &  &  & \\        
$\delta_{ov}$ 0.4 &  &  &  & \\
$f_{\rm wind}$  1    & 36.335 & 53.797 & 52.095 & 1.702  \\    
$f_{\rm wind}$  1.5  & 24.508 & 32.205 & 30.090 & 2.115  \\    
$f_{\rm wind}$  2    & 20.183 & 22.135 & 19.744 & 2.391  \\
&  &  &  & \\ 
$\delta_{ov}$ 0.5 &  &  &  & \\
$f_{\rm wind}$  1    & 30.383 & 48.080 & 46.399 & 1.682  \\    
$f_{\rm wind}$  1.5  & 22.559 & 29.984 & 28.017 & 1.968  \\    
$f_{\rm wind}$  2    & 19.189 & 20.896 & 18.737 & 2.159  \\
&  &  &  & \\    
\hline
\hline
\end{tabular}
\begin{tabular}{|lcccc|}
\hline\hline
model         & $R_{Max}$ & $M_{TAMS}$ & $M_{core,TAMS}$ & $M_{env,TAMS}$\\
              & [$\rsun$] & [$\msun$] & [$\msun$] & [$\msun$] \\
\hline\hline
&  &  &  & \\
$\delta_{ov}$ 0.12 &  &  &  & \\
$f_{\rm wind}$  1    & $>$2000 & 86.568 & 58.490 & 28.078  \\    
$f_{\rm wind}$  1.5  & 40.419 & 62.672 & 52.644 & 10.028  \\    
$f_{\rm wind}$  2    & 25.070 & 32.530 & 30.080 & 2.451  \\ 
&  &  &  & \\    
$\delta_{ov}$ 0.16 &  &  &  & \\
$f_{\rm wind}$  1    & $>$2000 & 85.196 & 60.295 & 24.900  \\    
$f_{\rm wind}$  1.5  & 33.693 & 49.404 & 47.764 & 1.640  \\    
$f_{\rm wind}$  2    & 23.654 & 29.393 & 26.787 & 2.606  \\ 
&  &  &  & \\    
$\delta_{ov}$ 0.2 &  &  &  & \\
$f_{\rm wind}$  1    & $>$2000 & 84.293 & 61.686 & 22.607  \\    
$f_{\rm wind}$  1.5  & 29.989 & 42.995 & 40.967 & 2.029  \\    
$f_{\rm wind}$  2    & 22.569 & 27.366 & 24.724 & 2.642  \\ 
&  &  &  & \\    
$\delta_{ov}$ 0.33 &  &  &  & \\
$f_{\rm wind}$  1    & 41.232 & 59.079 & 57.538 & 1.541 \\    
$f_{\rm wind}$  1.5  & 24.696 & 34.222 & 31.988 & 2.234  \\    
$f_{\rm wind}$  2    & 20.022 & 23.455 & 20.948 & 2.507  \\ 
&  &  &  & \\    
$\delta_{ov}$ 0.4 &  &  &  & \\
$f_{\rm wind}$  1    & 33.053 & 51.420 & 49.583 & 1.838 \\    
$f_{\rm wind}$  1.5  & 23.301 & 32.049 & 29.895 & 2.154 \\    
$f_{\rm wind}$  2    & 19.042 & 22.120 & 19.747 & 2.373 \\
&  &  &  & \\    
$\delta_{ov}$ 0.5 &  &  &  & \\
$f_{\rm wind}$  1    & 27.484 & 45.711 & 43.895 & 1.816  \\    
$f_{\rm wind}$  1.5  & 21.311 & 29.517 & 27.527 & 1.990  \\    
$f_{\rm wind}$  2    & 17.986 & 20.735 & 18.555 & 2.181  \\
&  &  &  & \\    
\hline
\hline
\end{tabular}
\end{table}

\end{appendices}

\end{document}